\documentclass[12pt]{article}
\usepackage{graphicx}

\newcommand\pubnumber{SNSN-323-63}
\newcommand\pubdate{\today}

\def\institute{Institute of High Energy Physics (IHEP), Beijing, China}
\def\support{\footnote{On behalf of the CMS Collaboration.}}

\def\Title#1{\begin{center} {\Large #1 } \end{center}}
\def\Author#1{\begin{center}{ \sc #1} \end{center}}
\def\Address#1{\begin{center}{ \it #1} \end{center}}

\newcommand\pubblock{\rightline{\begin{tabular}{l} \pubnumber\\
         \pubdate  \end{tabular}}}
\newenvironment{Abstract}{\begin{quotation}  }{\end{quotation}}
\newenvironment{Presented}{\begin{quotation} \begin{center} 
             PRESENTED AT\end{center}\bigskip 
      \begin{center}\begin{large}}{\end{large}\end{center} \end{quotation}}

\usepackage{xspace}          
\usepackage{wrapfig}   
\usepackage{amsmath}
\usepackage{xcolor}
\usepackage{bm}
\usepackage{bbm}
\usepackage{hyperref}
\newcommand{\ttbar}{\ensuremath{\mathrm{t}\bar{\mathrm{t}}}\xspace}
\newcommand{\ttZ}{\ensuremath{\mathrm{t}\bar{\mathrm{t}}\textrm{Z}}\xspace}
\newcommand{\ttX}{\ensuremath{\mathrm{t}\bar{\mathrm{t}}\textrm{X}}\xspace}
\newcommand{\ptvecmiss}{\ensuremath{{\vec p}_{\mathrm{T}}^{\kern3pt\mathrm{miss}}}\xspace}
\newcommand{\mtw}{\ensuremath{m_{\textrm{T}}(\textrm{W})}\xspace}
\newcommand{\POWHEG} {{\textsc{powheg}}\xspace}
\newcommand{\MADGRAPH} {\textsc{MadGraph}\xspace}
\newcommand{\MCATNLO} {\textsc{mc@nlo}\xspace}
\newcommand{\MGvATNLO}{\MADGRAPH{}\textrm{5\_a}\MCATNLO\xspace}

\def\beq{\begin{equation}}
\def\eeq#1{\label{#1}\end{equation}}
\def\eeqn{\end{equation}}

\def\beqa{\begin{eqnarray}}
\def\eeqa#1{\label{#1}\end{eqnarray}}
\def\eeqan{\end{eqnarray}}

\let\bar=\overbar

\def\Dslash{\not{\hbox{\kern-4pt $D$}}}
\def\dslash{\not{\hbox{\kern-2pt $\del$}}}

\def\msb{{\bar{\ssstyle M \kern -1pt S}}}

\begin{document}
\begin{titlepage}
\pubblock

\vfill
\Title{Studies of top quark spin and polarization in CMS}
\vfill
\Author{ Fabio Iemmi\support}
\Address{\institute}
\vfill
\begin{Abstract}
We describe the latest results obtained by the CMS Collaboration on top quark spin and polarization properties. The top quark spin asymmetry is measured both targeting single-top quark production in the $t$-channel and single-top quark production in association with a Z boson. Additionally, all the independent coefficients of the spin-dependent part of the top quark-antiquark production density matrix are measured and the results are extrapolated to the High-Luminosity LHC scenario.
\end{Abstract}
\vfill
\begin{Presented}
$15^\mathrm{th}$ International Workshop on Top Quark Physics\\
Durham, UK, 4--9 September, 2022
\end{Presented}
\vfill
\end{titlepage}
\def\thefootnote{\fnsymbol{footnote}}
\setcounter{footnote}{0}

\section{Introduction}

The top quark (t) is the heaviest fermion in the standard model (SM), its mass being comparable to the mass of atoms such as tungsten. This has important consequences on its properties, the most quoted of all being the fact that top quarks hadronize before forming any bound state. Additionally, top quarks decay before the strong interaction described by quantum chromodynamics (QCD) can randomize their spins. 
At the CERN Large Hadron Collider (LHC), the main production mechanism for top quarks is the associated production of a quark-antiquark pair, \ttbar, which is induced by QCD interactions. Thus, top quarks produced in pairs have no single spin configuration in any basis. On the other hand, production mechanisms induced by pure electroweak processes, such as the production of a single top quark in the $t$-channel, lead to strongly polarized top quarks, due to the V--A nature of the interaction happening at the Wtb vertex.

The polarization properties of the top quark are usually studied analyzing its decay products. In the case of a leptonic decay of a top quark produced in the $t$-channel, a powerful observable is the cosine of the polarization angle, $\cos \theta^*_{\textrm{pol}}$, defined as
\begin{equation}
\cos \theta^*_{\text{pol}} = \frac{\vec{p}^*_{q^{\prime}} \cdot \vec{p}^*_{\ell}}{|\vec{p}^*_{q^{\prime}}| |\vec{p}^*_{\ell}| },
\end{equation}
where $\ell$ is the lepton coming from the W boson decay, $q^\prime$ is the spectator quark and the superscript $^*$ indicates that the quantities are computed in the top quark rest frame. 

In addition, it holds true that the top quark spin asymmetry $A_\ell$ is related to the normalized, differential cross section as a function of $\cos \theta^*_{\textrm{pol}}$ by the relation
\begin{equation}
\frac{1}{\sigma}\frac{\text{d} \sigma}{\text{d}\cos \theta^*_{\text{pol}}} = \frac{1}{2} \left(1 + 2A_\ell \cos \theta^*_{\text{pol}} \right),
\label{eq:diff}
\end{equation}
so that it is possible to infer the top quark spin asymmetry by measuring the aforementioned differential cross section.

\section{Top quark spin asymmetry in \textit{t}-channel single-top production}
The CMS Collaboration \cite{bib:CMS} performed a measurement \cite{bib:tchan} of the top quark spin asymmetry targeting the production of a single top quark in the $t$-channel. This measurement uses the dataset collected in the 2016 data taking period, corresponding to $35.9\, \textrm{fb}^{-1}$, and targets the leptonic decay of the top quark. Thus, events containing one isolated muon or electron and two or three jets are analyzed. Events are split based on the jet and b tagged jet multiplicities to form three $N\textrm{j}M\textrm{b}$ categories: 2j1b, which is signal enriched; 3j2b, which is enriched in the main background of the analysis, namely \ttbar; and 2j0b, which is used for validation purposes.

One top quark candidate per event is reconstructed in the 2j1b category assuming $t$-channel single top production. As a first step, the component of the neutrino candidate momentum along the beam axis is determined by imposing a W boson mass constraint to the system formed by the charged lepton and the missing transverse momentum \ptvecmiss, the latter being interpreted as the projection in the transverse plane of the four-momentum of the neutrino itself. Then, the top quark candidate is built starting from the four-momenta of the charged lepton, the neutrino, and the b tagged jet.

Since the Monte Carlo (MC) simulations of the QCD background are found to be not reliable, this background is estimated with a procedure based on the definition of a sideband region in data. 
First, templates of the distribution for the transverse mass of the W boson, \mtw, from multijet events are obtained from data in the sideband region. Then, their normalizations are estimated in a second step through a template-based maximum-likelihood (ML) fit to the events in the 2j1b and 3j2b categories, simultaneously with the number of signal events, as described in the following paragraphs.

The number of $t$-channel single top quark events in data is determined from a ML fit to the \mtw distribution and to two BDT discriminant distributions. The first BDT, labeled $\textrm{BDT}_{t\textrm{-ch}}$, is trained to separate $t$-channel events from other background events, 
while the second BDT, labeled $\textrm{BDT}_{\ttbar/\textrm{W}}$, is trained to separate \ttbar events from W+jets events. The final fit is performed on the  $\textrm{BDT}_{t\textrm{-ch}}$, $\textrm{BDT}_{\ttbar/\textrm{W}}$ and \mtw distributions in the 2j1b category and on the \mtw distribution in the 3j2b category. 
To perform differential measurements, each observable of interest is divided in bins and separate fits are performed in each bin. The results are subsequently unfolded to the parton level.

\begin{figure}[]
\centering
\includegraphics[width=0.5\columnwidth]{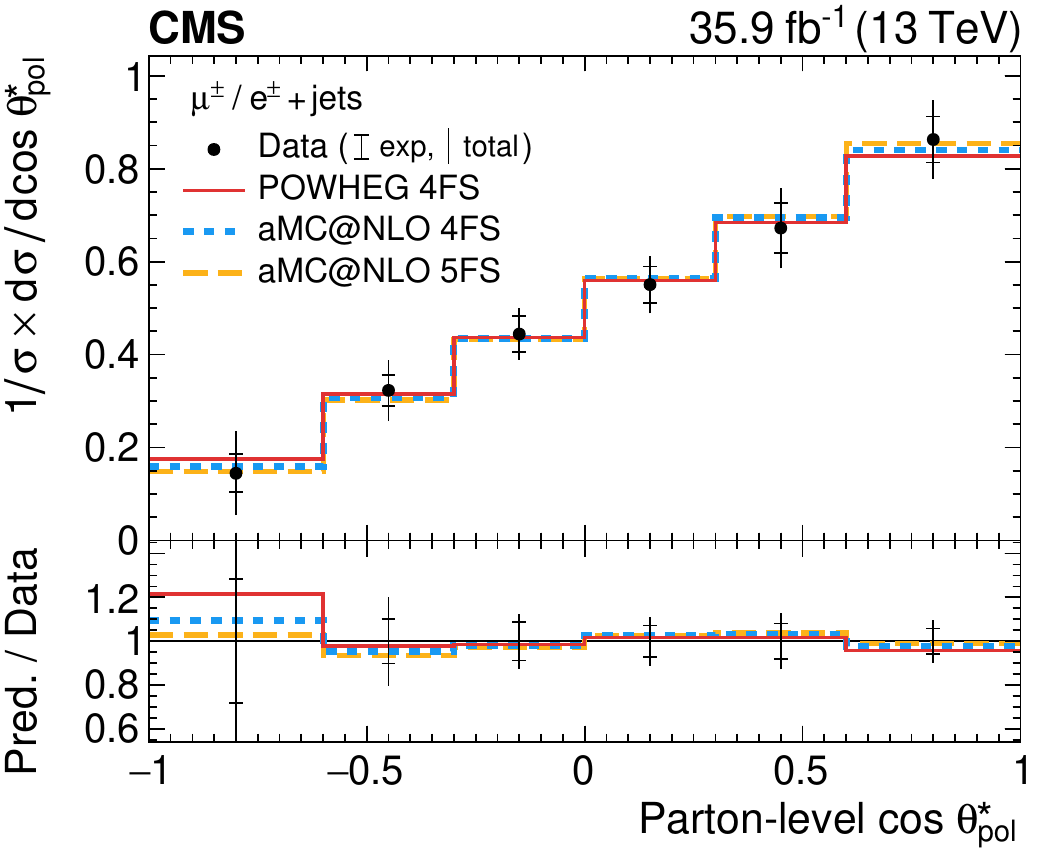}
\caption{Normalized, differential cross section for $t$-channel single top events as a function of $\cos \theta^*_{\textrm{pol}}$.}
\label{fig:diffsigma}
\end{figure}

The normalized differential cross section as a function of the cosine of the polarization angle is shown in Fig. \ref{fig:diffsigma}. By performing a $\chi^2$-based fit based on the functional dependence discussed in Eq. \ref{eq:diff}, the top quark spin asymmetry is found to be

\begin{equation}
A_\ell = 0.440 \pm 0.031 (\textrm{stat+exp}) \pm 0.062 (\textrm{theo}),
\end{equation}
which is in good agreement with the SM prediction of 0.436 obtained with \POWHEG with comparably negligible uncertainty. This measurement is dominated by theory uncertainties, the most relevant being the ones related to the top quark mass and to the \ttbar parton showering.

\section{Top quark spin asymmetry in tZq production}
The CMS Collaboration also performed a measurement \cite{bib:tzq} of the top quark spin asymmetry targeting the production of a single top quark in association with a Z boson (tZq).
This measurement uses the full Run2 dataset corresponding to  $139\, \textrm{fb}^{-1}$ and targets the leptonic decays of the top quark and of the Z boson. Thus, the analysis selects events with exactly three leptons (electrons or muons) containing an opposite-sign-same-flavor lepton pair. Similarly to what it has been described before, events are split in $N\textrm{j}M\textrm{b}$ categories based on the jet and b tagged jet multiplicities, with the category having $1 < N < 4$ and $M\geq 1$ being signal enriched. Such category is complemented by several control and validation regions to constrain the systematic uncertainties and validate the main steps of the analysis.

The top quark candidate in the event is reconstructed with a technique similar to the one described for the previous measurement. 

The main backgrounds of this analysis can be split into two categories: backgrounds containing prompt leptons from the hard interaction, and backgrounds containing non-prompt leptons, coming from the decay of heavy-flavored hadrons or from misidentified objects. While the former are estimated using simulation, the latter are estimated from data using a ``fake rate'' technique.
First, the probability for a non-prompt lepton to pass the analysis cuts (fake rate) is estimated in data from a sample enriched in QCD events. Then, this probability is applied to events in a region containing non-prompt lepton that do not pass the analysis cuts to obtain the expected yield from non-prompt backgrounds in the signal region.

The signal extraction relies on a multiclass neural network, trained to separate the tZq signal from the \ttZ, WZ, \ttX and minor backgrounds. After a full validation of the input features is performed, the outputs of the tZq and \ttZ output nodes are used in a ML fit to extract the signal, together with \mtw distributions and counting experiments in several control regions. To perform differential measurements, each observable of interest is divided in bins and separate fits are performed in each bin. The results are subsequently unfolded to the parton level.

\begin{figure}[]
\centering
\includegraphics[width=0.5\columnwidth]{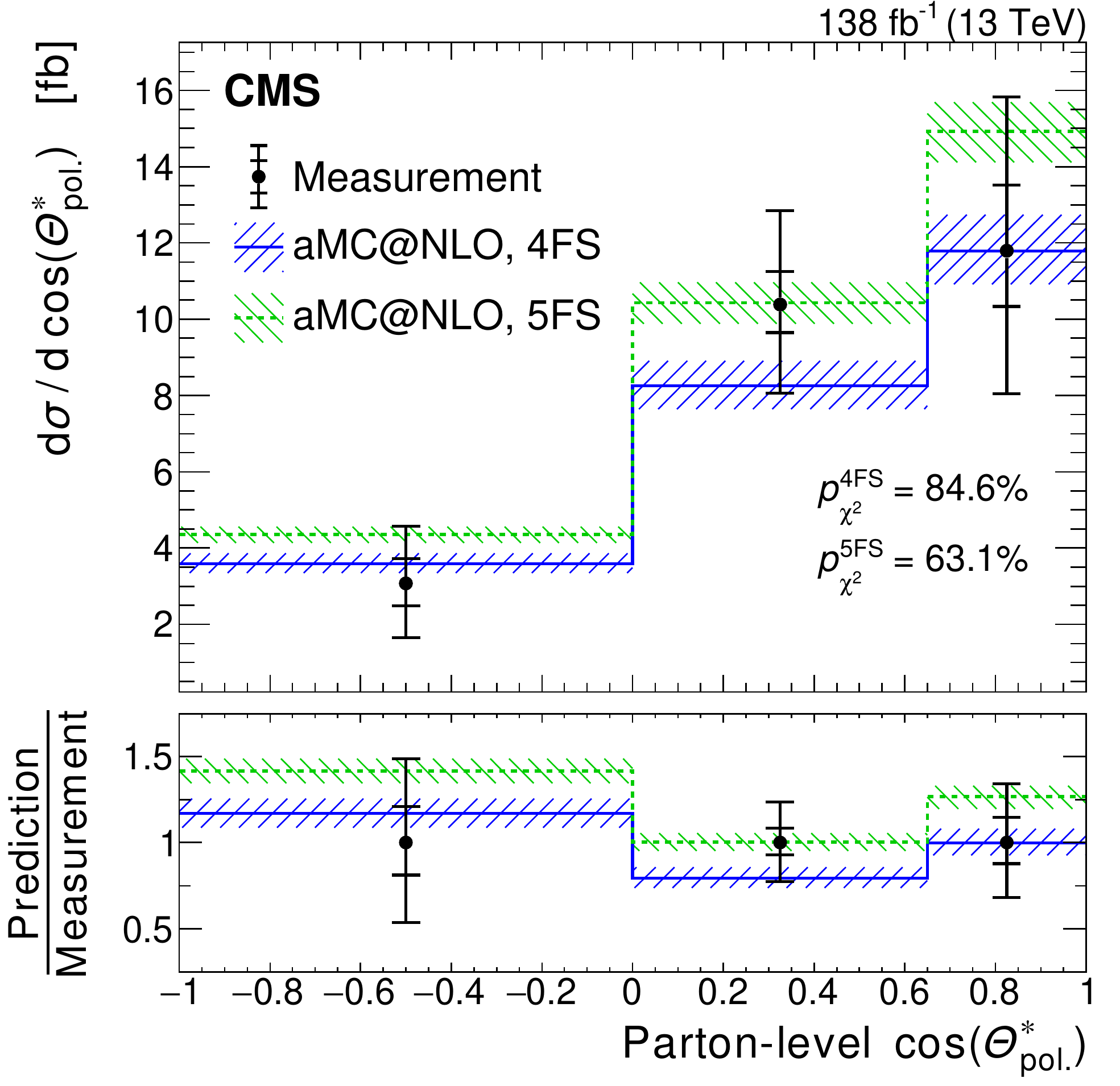}
\caption{Normalized, differential cross section for tZq events as a function of $\cos \theta^*_{\textrm{pol}}$.}
\label{fig:diffsigma2}
\end{figure}

Similarly to the previously-described measurement, the normalized differential cross section as a function of the cosine of the polarization angle shown in Fig. \ref{fig:diffsigma2} is used to extract the top quark spin asymmetry, which is found to be

\begin{equation}
A_\ell = 0.54 \pm 0.16 (\textrm{stat}) \pm 0.06 (\textrm{syst}),
\end{equation}
which is in good agreement with the SM prediction of 0.44 (0.45) obtained with \MGvATNLO in the 4-flavor (5-flavor) scheme with comparably negligible uncertainty. In constrast to the previous one, this measurement is dominated by statistical uncertainties, the tZq production mode being far rarer than the $t$-channel single top production.

\section{Polarization and spin coefficients in $\textrm{t}\bar{\textrm{t}}$ events}
Top quarks are mainly produced in \ttbar pairs at the LHC, such pairs being unpolarized at tree level due to parity-conserving and time-invariant properties of QCD, which is the process initiating the production. Nevertheless, small sources of polarization can come from electroweak corrections and absorbtive terms appearing at one-loop level. The CMS Collaboration performed an analysis \cite{bib:tt} whose goal is to measure all the 15 independent coefficients of the spin-dependent part of the \ttbar production density matrix, called $\mathcal{R}$ in the following.
Similarly to what we discussed in the introduction, a relation holds between the following normalized double differential \ttbar cross section and the aforementioned coefficients,

\begin{equation}
\frac{1}{\sigma}\frac{\text{d}^2 \sigma}{\text{d}\cos \theta^i_{1} \text{d}\cos \theta^j_{2}} = \frac{1}{4} \left( 1 + B^i_1 \cos \theta^i_{1} + B^j_2 \cos \theta^j_{2} - C_{ij} \cos \theta^i_{1} \cos \theta^j_{2} \right),
\end{equation}
where $\theta^i_{1}$ ($\theta^j_{2}$) is the angle of the positively (negatively) charged lepton with respect to axis $i$ ($j$) in the rest frame of the t ($\bar{\textrm{t}}$) quark, while $\boldsymbol{\tilde{B}}^{\pm}$ are three-dimensional vectors describing the top quark and antiquark polarization in each direction, and $\tilde{C}$ is a 3x3 matrix of coefficients describing spin correlations between the top quark and the top antiquark. Integrating angles out one at a time, 15 single-differential cross sections can be obtained and used to measure the coefficients of the $\mathcal{R}$ matrix.

The analysis uses $35.9\, \textrm{fb}^{-1}$ of data and selects events with exactly two leptons with opposite charge, at least two jets and at least one b tagged jet. The main backgrounds of this analysis are other \ttbar decay channels, tW production and the production of a Z boson with additional jets. The \ttbar pair is reconstructed with a kinematic fit constrained to the W boson and top quark masses and, after the angular observables of interest are built, the differential spectra are computed and unfolded at parton level. 

\begin{figure}[ht]
\centering
\includegraphics[width=0.40\columnwidth]{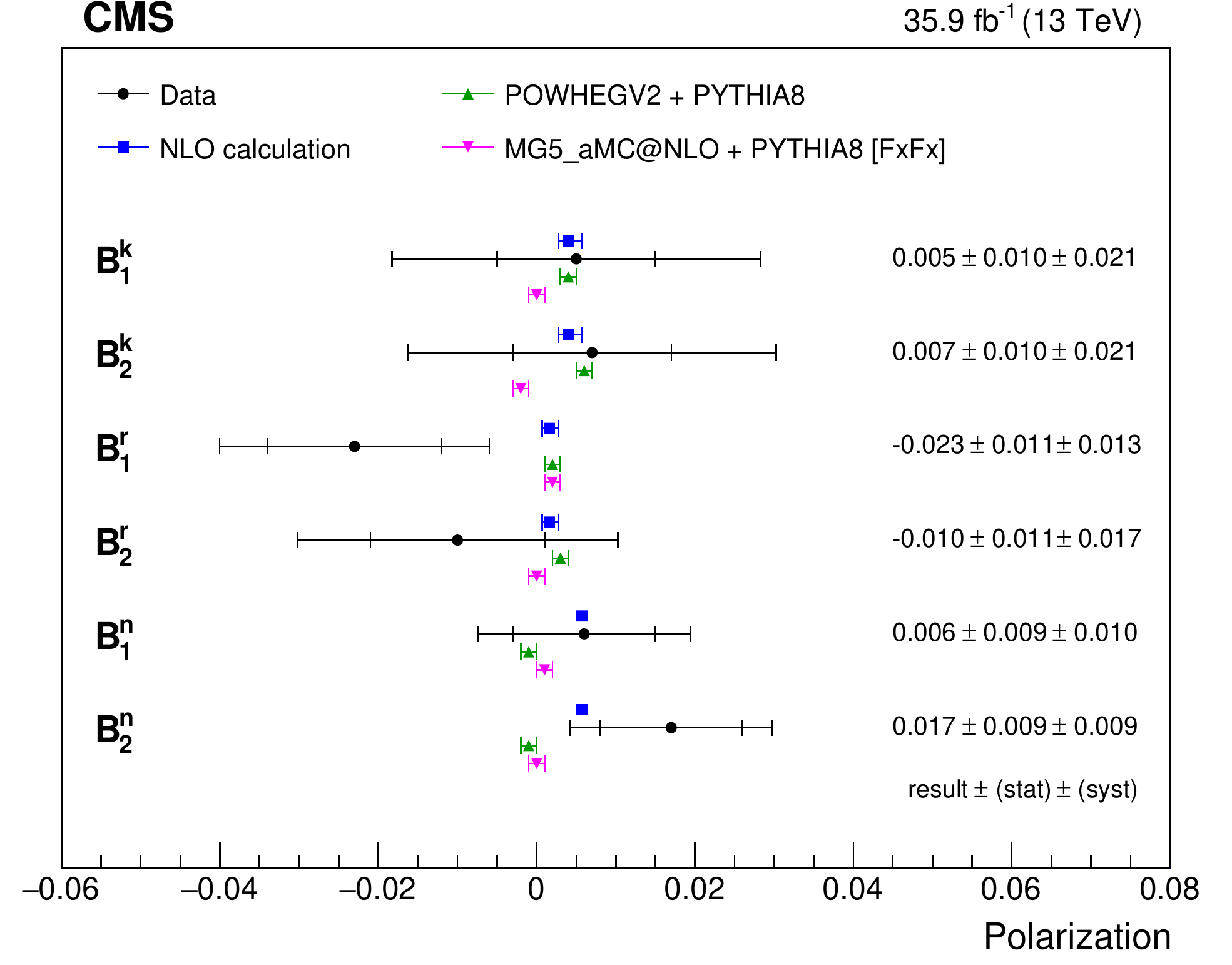}
\includegraphics[width=0.40\columnwidth]{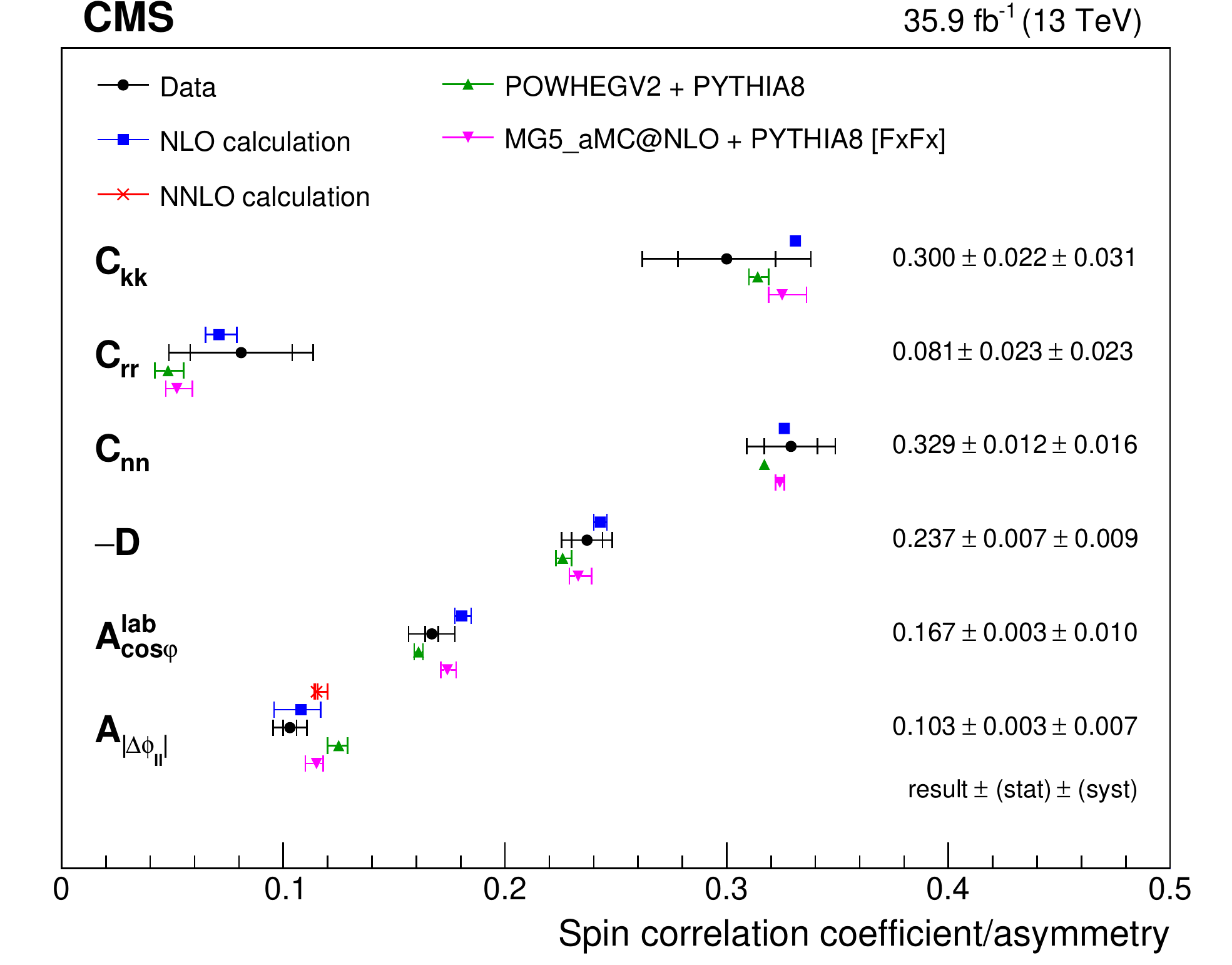}
\includegraphics[width=0.40\columnwidth]{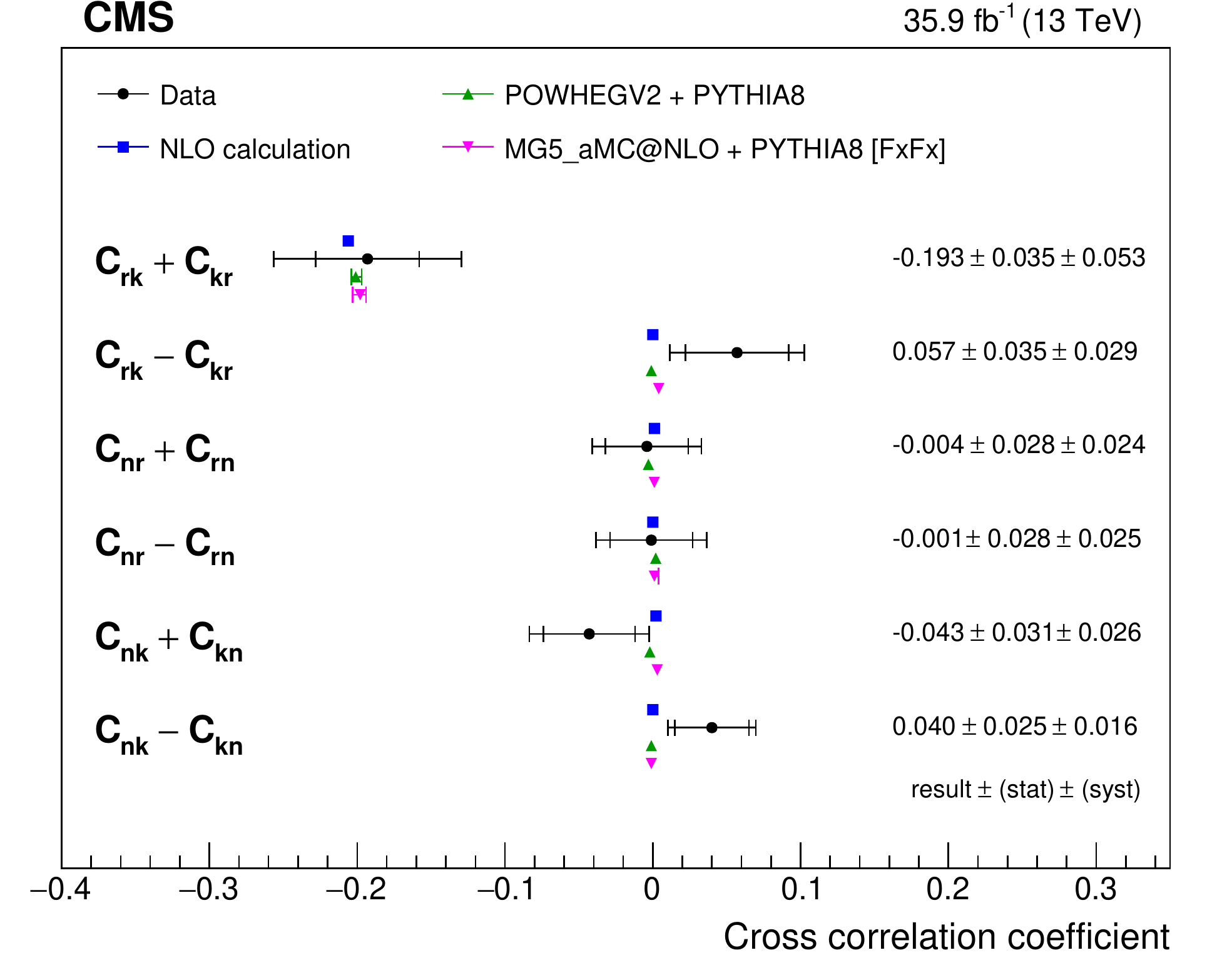}
\caption{Measured and expected values of the independent coefficients of the spin-dependent part of the \ttbar production density matrix.}
\label{fig:diffttbar}
\end{figure}

In general, a good agreement is found between the SM prediction and the measured values for all the 15 coefficients mentioned above, the impact of statistical and systematic uncertainties being comparable. These results are summarized in Fig. \ref{fig:diffttbar}.

The same analysis setup is also used to project this measurement to the High-Luminosity LHC scenario \cite{bib:ttftr}. By assuming trigger and detector performances comparable to the ones in Run2, and a better control of experimental and theoretical uncertainties, an unprecedented precision of up to 3\% on the spin correlation observables is obtained.

\section{Conclusions}
The CMS Collaboration has performed several measurements of the top quark spin and polarization properties. The top quark spin asymmetry has been measured in $t$-channel and tZq events. The independent coefficients of the spin-dependent part of the \ttbar production density matrix have been measured as well. All the aforementioned results are found to be in agreement with the SM predictions. The results on the spin coefficients have also been projected to the High-Luminosity LHC scenario, demonstrating that unprecedented precision can be achieved.

\end{document}